%% file: manuscript.tex
\documentclass[preprint,11pt]{elsarticle}
\journal{Information and Software Technology}
\journal{Journal of Systems and Software}
\usepackage{amsmath,amssymb,amsfonts}
\usepackage{algorithmic}
\usepackage{subcaption}
\usepackage{graphicx}
\usepackage{textcomp}
\usepackage{xcolor}
\usepackage{multirow}
\usepackage{tabularx}
\usepackage{flushend}
\usepackage{array}
\newcolumntype{?}{!{\vrule width 1pt}}
\usepackage{float}
\newcolumntype{C}[1]{>{\centering\arraybackslash}p{#1}}
\usepackage{todonotes}
\def\BibTeX{{\rm B\kern-.05em{\sc i\kern-.025em b}\kern-.08em
    T\kern-.1667em\lower.7ex\hbox{E}\kern-.125emX}}
\usepackage{tikz}
\tikzstyle{mybox} = [draw=black, very thick, rectangle, rounded corners, inner ysep=5pt, inner xsep=5pt]

\begin{document}
\begin{frontmatter}
\title{Some SonarQube Issues have a Significant but Small Effect on Faults and Changes. A large-scale empirical study
}

\author {Valentina Lenarduzzi}
\ead{valentina.lenarduzzi@tuni.fi}

\author {Nyyti Saarim\"{a}ki}
\ead{nyyti.saarimaki@tuni.fi}

\author {Davide Taibi}
\ead{davide.taibi@tuni.fi}

\address {Tampere University, Tampere (Finland)}

\begin{abstract}
\textit{Context}. Companies commonly 
 invest effort to 
remove technical issues believed to impact software qualities, such as removing anti-patterns or coding styles violations. \\
\textit{Objective}. Our aim is to analyze the diffuseness of Technical Debt (TD) items in software systems and to assess their impact on code changes and fault-proneness, considering also the type of TD items and their severity.\\
\textit{Method}. We conducted a case study among 33 Java  projects from the  Apache Software Foundation (ASF) repository. We analyzed 726 commits containing 27K faults and 12M changes. The projects violated 173 SonarQube rules generating more than 95K TD items in more than 200K classes.\\
\textit{Results}. Clean classes (classes not affected by TD items) are less change-prone than dirty ones, but the difference between the groups is small. Clean classes are slightly more change-prone than classes affected by TD items of type Code Smell or Security Vulnerability. As for fault-proneness, there is no difference between clean and dirty classes. Moreover, we found a lot of incongruities in the type and severity level assigned by SonarQube.\\
\textit{Conclusions}. Our result can be useful for practitioners to understand which TD items they should refactor and for researchers to bridge the missing gaps. They can also support companies and tool vendors in identifying TD items as accurately as possible. 

\end{abstract}

\end{frontmatter}

\input{Sections/Introduction.tex}
\input{Sections/Background.tex}

\input{Sections/RelatedWork.tex}

\input{Sections/CaseStudy.tex}
\input{Sections/Results.tex}
\input{Sections/Discussion.tex}
\input{Sections/Threats.tex}

\input{Sections/Conclusion.tex}

\section*{References}
\bibliographystyle{model1-num-names}
\bibliography{manuscript}

\end{document}

%% file: Sections/Introduction.tex
\section{Introduction}
\label{Introduction}
Companies commonly spend time to improve the quality of the software they develop, investing effort into refactoring activities aimed at removing technical issues believed to impact software qualities. Technical issues include any kind of information that can be derived from the source code and from the software process, such as usage of specific patterns, compliance with coding or documentation conventions, architectural issues, and many others. If such issues are not fixed, they generate Technical Debt.

Technical Debt (TD) is a metaphor from the economic domain that refers to different software maintenance activities that are postponed in favor of the development of new features in order to get short-term payoff~\cite{Cunningham1992}. Just as in the case of financial debt, the additional cost will be paid later. The growth of TD commonly slows down the development process~\cite{Cunningham1992}\cite{Li2007}. 

Different types of TD exist: requirements debt, code debt, architectural debt, design debt, test debt, build debt, documentation debt, infrastructure debt, versioning debt, and defect debt~\cite{Li2007}.
Some types of TD, such as ''code TD'', can be measured using static analysis tools, which is why several companies have started to adopt code TD analysis tools such as SonarQube, Cast, and Coverity, investing a rather large amount of their budget into refactoring activities recommended by these tools. This is certainly a very encouraging sign of a software engineering research topic receiving balanced attention from both communities, research and industry. 

SonarQube is one of the most frequently used open-source code TD analysis tools~\cite{LenarduzziSEDA2019}, having been adopted by more than 85K organizations\footnote{https://www.sonarqube.org}, including nearly 15K public open-source projects\footnote{https://sonarcloud.io/explore/projects}. SonarQube allows code TD management by monitoring the evolution of TD and alerting developers if certain TD items increase beyond a specified threshold or, even worse, grow out of control. TD monitoring can also be used to support the prioritization of repayment actions where TD items are resolved (e.g., through refactoring)~\cite{Digkas2018}\cite{SaarimakiTechDebt2019}. 
SonarQube monitors the TD analyzing code compliance against a set of rules. If the code violates a rule, SonarQube adds the time needed to refactor the violated rule as part of the technical debt, thereby creating an issue. In this paper we refer to these issues with the term ''TD items''. 

SonarQube classifies TD items into three main categories: \textit{Code Smells}, i.e., TD items that increase change-proneness and the related maintenance effort;  \textit{Bugs}, i.e., TD items that will result in a fault; and \textit{Security Vulnerabilities}\footnote{\label{sq-rules}SonarQube Rules: https://docs.sonarqube.org/display/SONAR/Rules Last Access: May 2018}.

It is important to note that the term ''code smells'' adopted in SonarQube does not refer to the commonly known term code smells defined by Fowler et al.~\cite{Fowler1999}. 
SonarQube also classifies the rules into five \textit{severity} levels:  Blocker, Critical, Major, Minor, and Info.
The complete list of violations can be found in the replication package\footnote{Replication Package: https://figshare.com/s/240a036f163759b1ec97}.

Even if developers are not sure about the usefulness of the rules, they do pay attention to their categories and priorities and tend to remove violations related to rules with a high level of severity in order to avoid the potential risk of faults~\cite{Vassallo2018}\cite{Palomba2018}\cite{TaibiIST2017}. However, to the best of our knowledge, there are currently no studies that have investigated both the fault-proneness of rules classified as \textit{Bugs} and the change-proneness of rules classified as \textit{Code Smells}. 

Therefore, in order to help both practitioners and researchers understand whether SonarQube rules are actually fault- or change-prone, we designed and conducted an empirical study analyzing the evolution of 33 projects every six months. 
Our goal was to assess the impact of the TD items on change- and fault-proneness as well as considering the severity of this impact.  


The result of this work can benefit several groups. It helps practitioners to understand which TD items they should refactor and researchers to bridge the missing gaps, and supports companies and tool vendors in identifying TD items as accurately as possible. 

\textbf{Structure of the paper}. Section~\ref{Background} describes the basic concepts underlying this work, while Section~\ref{RW} presents some related work done by researchers in recent years. In  Section~\ref{CS}, we describe the design of our case study, defining the research questions, metrics, and hypotheses, and describing the study context with the data collection and data analysis protocol. In Section~\ref{Results}, we present the achieved results and discuss them in Section~\ref{Discussion}. In Section~\ref{Threats}, we identify the threats to the validity of our study, and in  Section~\ref{Conclusion}, we draw conclusions and give an outlook on possible future work. 

%% file: Sections/Background.tex
\section{Background}
\label{Background}
SonarQube is one of the most common open-source static code analysis tools for measuring code technical debt~\cite{Vassallo2018},\cite{LenarduzziSEDA2019}. SonarQube is provided as a service by the sonarcloud.io platform or can be downloaded and executed on a private server.

SonarQube calculates several metrics such as  number of lines of code and code complexity, and verifies the code's compliance against a specific set of "coding rules" defined for most common development languages. Moreover, it defines a set of thresholds ("quality gates") for each metric and rule. 
If the analyzed source code violates a coding rule, or if a metric is outside a predefined threshold (also named "gate"), SonarQube generates an issue (a "TD item"). The time needed to remove these issues (remediation effort) is used to calculate the remediation cost and the technical debt. SonarQube includes Reliability, Maintainability, and Security rules. Moreover, SonarQube claims that zero false positives are expected from the Reliability and Maintainability rules\footnote{SonarQube Rules:https://docs.sonarqube.org/display/SONAR/Rules}. 

Reliability rules, also named \textit{Bugs}, create issues that ''represent something wrong in the code'' and that will soon be reflected in a bug.  \textit{Code smells} are considered  ''maintainability-related issues'' in the code that decrease code readability and code modifiability. It is important to note that the term ''code smells'' adopted in SonarQube does not refer to the commonly known term code smells defined by Fowler et al.~\cite{Fowler1999}, but to a different set of rules. 

SonarQube also classifies the rules into five \textit{severity} levels\footnote{SonarQube Issues and Rules Severity:'  https://docs.sonarqube.org/display/SONAR/Issues}:
\begin{itemize}
\item \textit{BLOCKER}: ''Bug with a high probability to impact the behavior of the application in production: memory leak, unclosed JDBC connection.'' SonarQube recommends  immediately reviewing such an issue
\item \textit{CRITICAL}: ''Either a bug with a low probability to impact the behavior of the application in production or an issue which represents a security flaw: empty catch block, SQL injection'' SonarQube recommends immediately reviewing such an issue 
\item \textit{MAJOR}: ''Quality flaw which can highly impact the developer productivity: uncovered piece of code, duplicated blocks, unused parameters''
\item \textit{MINOR}: ''Quality flaw which can slightly impact the developer productivity: lines should not be too long, \"switch\" statements should have at least 3 cases, ...''
\item \textit{INFO}: ''Neither a bug nor a quality flaw, just a finding.''
\end{itemize}
The complete list of violations can be found in the online raw data (Section~\ref{Replicability}).

%% file: Sections/RelatedWork.tex
\section{Related Work}
\label{RW}
In this Section, we report the most relevant works on the diffuseness, change- and fault-proneness of code TD items.
\subsection{Diffuseness of Technical Debt issues}
To the best of our knowledge, the vast majority of publications in this field investigate the distribution and evolution of code smells~\cite{Fowler1999} and antipatterns~\cite{brown1998}, but few papers investigated SonarQube violations.

Vaucher et al.~\cite{Vaucher2009} considered God Class code smells in their study, focusing on whether these affect software systems for long periods of time and making a comparison with whether the code smell is refactored.

Olbrich et al.~\cite{Olbrich2009} investigated the evolution of two code smells, God Class and Shotgun Surgery. They found that the distribution over time of these code smells is not constant; they increase during some periods and decrease in others, without any correlation with project size. 

In contrast, Chatzigeorgiou and Manakos~\cite{Chatzigeorgiou2010} investigated the evolution of several code smells and found that the number of instances of code smells increases constantly over time. This was also confirmed by Arcoverde et al.~\cite{Arcoverde2011}, who analyzed the longevity of code smells.

Tufano et al.~\cite{Tufano2017} showed that close to 80\% of the code smells are never removed from the code, and that those code smells that are removed are eliminated by removing the smelly artifact and not as a result of refactoring activities. 

Palomba et al.~\cite{Palomba2018} conducted a study on 395 versions of 30 different open-source Java applications, investigating the diffuseness of 13 code smells and their impact on two software qualities: change-  and fault-proneness. They analyzed 17,350 instances of 13 code smells, which were identified by applying a metric-based approach. 
Out of the 13 code smells, only seven were highly diffused smells; their removal would result in great benefit to the software in terms of change-proneness.
In contrast, the benefit regarding fault-proneness was very limited or non-existent. So programmers should keep an eye on these smells and do refactoring where needed in order to improve the overall maintainability of the code.

To the best of our knowledge, only four works consider code TD calculated by SonarQube ~\cite{SaarimakiTechDebt2019}\cite{Digkas2017}\cite{Digkas2018}\cite{AMANATIDIS2017}.

Saarim{\"a}ki et al.~\cite{SaarimakiTechDebt2019} investigated the diffuseness of TD items in Java projects, reporting that the most frequently introduced TD items are related to low-level coding issues. The authors did not consider the remediation time for TD.

Digkas et al.~\cite{Digkas2017} investigated the evolution of Technical Debt over a period of five years at the granularity level of weekly snapshots. They considered as context 66 open-source software projects from the Apache ecosystem.
Moreover, they characterized the lower-level constituent components of Technical Debt.
The results showed a significant increase in terms of size, number of issues, and complexity metrics of the analyzed projects.
However, they also discovered that normalized TD decreased as the aforementioned project metrics evolved.

Moreover, Digkas et al.~\cite{Digkas2018} investigated in a subsequent study how TD accumulates as a result of software maintenance activities. As context, they selected 57 open-source Java software projects from the Apache Software Foundation and analyzed them at the temporal granularity level of weekly snapshots, also focusing on the types of issues being fixed. The results showed that the largest percentage of Technical Debt repayment is created by a small subset of issue types.

Amanatidis et al.~\cite{AMANATIDIS2017} investigated the accumulation of TD in PHP applications (since a large portion of software applications are deployed on the web), focusing on the relation between debt amount and interest to be paid during corrective maintenance activities.
They analyzed ten open-source PHP projects from the perspective of corrective maintenance frequency and corrective maintenance effort related to interest amount and found a positive correlation between interest and the amount of accumulated TD. 

\subsection{Change- and Fault-proneness of Technical Debt issues} 
Only two works investigated the change- and fault-proneness of TD items analyzed by SonarQube~\cite{Falessi2017}\cite{Tollin2017}.

Falessi et al.~\cite{Falessi2017} studied the distribution of 16 metrics and 106 SonarQube violations in an industrial project. They applied a \textit{What-if} approach with the goal of investigating what could happen if a specific sq-violation had not been introduced in the code and if the number of faulty classes  decreases in case the violation is not introduced. They compared  four Machine Learning (ML) techniques (Bagging, BayesNet, J48, and Logistic Regression) on the project and then applied the same techniques to a modified version of the code, where they had manually removed sq-violations. Their results showed that 20\% of the faults could have been avoided if the code smells had been removed. 

Tollin et al. \cite{Tollin2017} used ML to predict the change-proneness of classes based on SonarQube violations and their evolution. They investigated whether Sonar Qube violations would lead to an increase in the number of changes (code churns) in subsequent commits. The study was applied to two different industrial projects, written in C\# and JavaScript. The authors compared the prediction accuracy of Decision Trees, Random Forest, and Naive Bayes. They report that classes affected by more sq-violations have greater change-proneness. However, they did not prioritize or classify the most change-prone sq-violations. 



Other works investigated the fault proneness of different types of code smells~\cite{Fowler1999}, such as MVC smells~\cite{Aniche2018}, testing smells~\cite{ Bavota2015}, or Android smells~\cite{Kessentini2017}.



To the best of our knowledge, our work is the first study that investigated and  ranked SonarQube violations considering both  their change- and fault-proneness on the same set of projects. Moreover, differently than previous works, our work is the first work analyzing the accuracy of the SonarQube TD items classification, including TD items types and severity.

%% file: Sections/CaseStudy.tex
\section{Case Study Design}
\label{CS}

We designed our empirical study as a case study based on the guidelines defined by Runeson and H\"{o}st~\cite{Runeson2009}.
In this Section, we will describe the case study design including the goal and the research questions, the study context, the data collection, and the data analysis procedure. 

\subsection{Goal and Research Questions}
\label{RQ}

The goal of this study was to analyze the diffuseness of TD items in software systems and to assess their impact on the change- and fault-proneness of the code, considering also the type of technical debt issues and their severity.

Accordingly, to meet our expectation, we formulated the goal as follows, using the Goal/Question/Metric (GQM) template~\cite{Basili1994}: 

\vspace{2mm}
\begin{tabular}
{@{}p{1.5cm}p{8cm}@{}}
\textit{Purpose} & Analyze \\
\textit{Object} &  technical debt issues \\
\textit{Quality} & with respect to their fault- and change-proneness\\
\textit{Viewpoint} & from the point of view of developers \\ 
\textit{Context} & in the context of Java projects\\
 & \\
\end{tabular}

Based on the defined goal, we derived the following Research Questions (RQs): 

\noindent\textbf{RQ1} Are classes affected by TD items more change- or fault-prone than non-affected ones? 

\noindent\textbf{RQ2}  Are classes affected by TD items classified by SonarQube as \textit{different types} more change- or fault-prone than non-affected ones? 

\noindent\textbf{RQ3} Are classes affected by TD items classified by SonarQube with \textit{different levels of severity} more change- or fault-prone than non-affected ones?

\noindent\textbf{RQ4} How \textit{good is the classification} of the SonarQube rules? 

\vspace{2mm}
\textbf{RQ1} aims at measuring the magnitude if the change- and fault-proneness of these classes. We considered the number of changes and the number of bug fixes. Our hypothesis was that classes affected by TD items, \textit{independent of their type and severity} are more change- or fault-prone than non-affected ones. 

\textbf{RQ2} and \textbf{RQ3} aim at determining how the rules are grouped between different values of type (RQ2) and severity (RQ3) and what the relative distribution of different levels of severity and different types is in the analyzed projects. No studies have investigated yet whether the rules classified as ''\textit{Bugs}'' or ''\textit{Code Smells}'' are fault- or change-prone, according to the SonarQube classification.

Based on the definition of SonarQube ''\textit{Bugs}'' and  ''\textit{Code Smells}'', we hypothesized that classes affected by ''\textit{Bugs}'' are more fault-prone and classes affected by ''\textit{Code Smells}'' are more change-prone. 

Moreover, SonarQube assumes that higher level of severity assigned to the different rules suggests more intensity in changes or faults.  Therefore, we aim at understanding whether the severity level increases together with their actual fault- or change-proneness, considering  within the same type  (''\textit{Bugs}'' or ''\textit{Code Smells}'') and across types. 

\textbf{RQ4} aims at combining RQ2 and RQ3 to understand an eventual disagreement in the classification of SonarQube rules, considering both the type and severity of TD items. 
Therefore, we hypothesized that classes affected by ''\textit{Bugs}''
with a higher level of severity are more fault-prone than those affected by ''\textit{Bugs}'' with a lower level of severity or those not affected. 
In addition, for ''\textit{Bug}'', we hypothesized that classes affected by  ''\textit{Code Smells}'' with a higher level of severity are more change-prone than those with a lower level of severity ''\textit{Code Smells}'' or those not affected. 



\subsection{Context}
\label{Context}
For this study, we selected projects based on "criterion sampling"\cite{Patton2002}. 
The selected projects had to fulfill all of the following criteria: 

\begin{itemize}
\item Developed in Java
\item Older than three years
\item More than 500 commits
\item More than 100 classes
\item Usage of an issue tracking system with at least 100 issues reported 
\end{itemize}

Moreover, as recommended by Nagappan et al.~\cite{Nagappan2013}, we also tried to maximize diversity and representativeness by considering a comparable number of projects with respect to project age, size, and domain.
 
Based on these criteria, we selected 33 Java projects from the Apache Software Foundation (ASF) repository\footnote{http://apache.org}. This repository includes some of the most widely used software solutions. The available projects can be considered industrial and mature, due to the strict review and inclusion process required by the ASF. Moreover, the included projects have to keep on reviewing their code and follow a strict quality process\footnote{https://incubator.apache.org/policy/process.html}. 

We selected a comparable number of projects with respect to their domain, project age, size, and domain. 
Moreover, the  projects had be  older than three years, have more than 500 commits and 100 classes and must report at least 100 issues in Jira.


In Table~\ref{tab:SelectedProjects}, we report the list of the 33 projects we considered together with the number of analyzed commits, the project sizes (LOC) of the last analyzed commits, and the number of faults and changes in the commits.

\begin{table*}
\scriptsize
\centering
\caption{Description of the selected projects} 
\label{tab:SelectedProjects} 
\begin{tabular}
{@{}p{3.2cm}|p{0.5cm}|p{2.4cm}|p{1.2cm}|p{1cm}|p{1cm}|p{1.5cm}@{}}
\hline 
\textbf{Project Name} & \multicolumn{2}{c|}{\textbf{Analyzed Commits}} & \textbf{Last Commit LOC} & \textbf{Last Commit Classes} & \textbf{\# Faults} & \textbf{\# Changes}   \\ \cline{2-3}
&  \textbf{\# } & \textbf{Timeframe} &  & &   \\ \hline
Accumulo & 3 & 2011/10 - 2013/03 & 307,167 & 4,137 & 9,606 & 850,127 \\ 
Ambari & 8 & 2011/08 - 2015/08 & 774,181 & 3,047 & 7,110 & 677,251 \\ 
Atlas & 7 & 2014/11 - 2018/05 & 206,253 & 1,443 & 1,093  &  570,278 \\ 
Aurora & 16 & 2010/04 - 2018/03 & 103,395 & 1,028 & 19 & 485,132 \\ 
Batik & 3 & 2000/10 - 2002/04 & 141,990 & 1,969 & 54 & 365,951 \\ 
Beam & 3 & 2014/12 - 2016/06 & 135,199 & 2,421 & 51 & 616,983 \\ 
Cocoon & 7 & 2003/02 - 2006/08 & 398,984 & 3,120 & 227 & 2,546,947 \\ 
Commons BCEL & 32 & 2001/10 - 2018/02 & 43,803 & 522 & 129 & 589,220 \\ 
Commons BeanUtils & 33 & 2001/03 - 2018/06 & 35,769 & 332 & 1 & 448,335 \\ 
Commons CLI & 29 & 2002/06 - 2017/09 & 9,547 & 58 & 25 & 165,252 \\ 
Commons Codec & 30 & 2003/04 - 2018/02 & 21,932 & 147 & 111 & 125,920 \\ 
Commons Collections & 35 & 2001/04 - 2018/07 & 66,381 & 750 & 88 & 952,459 \\ 
Commons Configuration & 29 & 2003/12 - 2018/04 & 87,553 & 565 & 29 & 628,170 \\ 
Commons Daemon & 27 & 2003/09 - 2017/12 & 4,613 & 24 & 4 & 7,831 \\ 
Commons DBCP & 33 & 2001/04 - 2018/01 & 23,646 & 139 & 114 & 184,041 \\ 
Commons DbUtils & 26 & 2003/11 - 2018/02 & 8,441 & 108 & 17 & 40,708 \\ 
Commons Digester & 30 & 2001/05 - 2017/08 & 26,637 & 340 & 44 & 321,956 \\ 
Commons Exec & 21 & 2005/07 - 2017/11 & 4,815 & 56 & 40 & 21,020 \\ 
Commons FileUpload & 28 & 2002/03 - 2017/12 & 6,296 & 69 & 37 & 42,441 \\ 
Commons IO & 33 & 2002/01 - 2018/05 & 33,040 & 274 & 336 & 225,560\\ 
Commons Jelly & 24 & 2002/02 - 2017/05 & 30,100 & 584 & 29 & 205,691 \\ 
Commons JEXL & 31 & 2002/04 - 2018/02 & 27,821 & 333 & 180 & 187,596 \\ 
Commons JXPath & 29 & 2001/08 - 2017/11 & 28,688 & 253 & 30 & 188,336 \\ 
Commons Net & 32 & 2002/04 - 2018/01 & 30,956 & 276 & 114 & 428,427 \\ 
Commons OGNL & 8 & 2011/05 - 2016/10 & 22,567 & 333 & 1 & 39,623 \\ 
Commons Validator & 30 &  2002/01 - 2018/04 & 19,958 & 161 & 60 & 123,923 \\ 
Commons VFS & 32 & 2002/07 - 2018/04 & 32,400 & 432 & 152 & 453,798\\ 
Felix & 2 & 2005/07 - 2006/07 & 55,298 & 687 & 5,424 & 173,353 \\ 
HttpComponents Client & 25 & 2005/12 - 2018/04 & 74,396 & 779 & 15 & 853,118 \\ 
HttpComponents Core & 21 & 2005/02 - 2017/06 & 60,565 & 739 & 128 & 932,735 \\ 
MINA SSHD & 19 & 2008/12 - 2018/04 & 94,442 & 1,103 & 1,588 & 380,911 \\ 
Santuario Java & 33 & 2001/09 - 2018/01 & 124,782 & 839 & 99 & 602,433 \\ 
ZooKeeper & 7 & 2014/07 - 2018/01 & 72,223 & 835 & 385 & 35,846 \\

\hline
\textbf{Sum} & \textbf{726} &  & \textbf{2,528,636} & \textbf{27,903} & \textbf{27,340} & \textbf{12,373,716} \\ 
\hline
\end{tabular}
\end{table*}

\subsection{Data Collection}
\label{DataCollection}

All selected projects were cloned from their Git repositories. Each commit was analyzed for TD items using SonarQube. We used SonarQube's default rule set. 
We exported SonarQube violations as a CSV file using SonarQube APIs. The data is available in the replication package (Section~\ref{Replicability}). 

To calculate fault-proneness, we determined fault-inducing and bug-fixing commits from the projects' Git history. This was done using the SZZ algorithm, which is based on Git's annotate/blame feature~\cite{SZZ}. The algorithm has four steps. The first step fetches the issues from a bug tracking system. All of the projects analyzed in this paper use Jira as their bug tracking system. The second step preprocesses the git log output, and the third identifies the bug-fixing commits. This is possible because the AFS policies require developers to report the fault-ID in the commit message of a fault-fixing commit. Finally, the last step identifies the fault-introducing commits using the data gathered in the previous steps.

The analysis was performed by taking a snapshot of the main branch of each project every 180 days. The number of used commits varied between the projects. Table~\ref{tab:SelectedProjects} reports for each project the number of commits and the time frames the commits are taken from.

We selected 6-months snapshots since the changes between subsequent commits usually affect only a fraction of the classes and  the analysis of all the commits would have caused change- and fault-proneness to be zero for almost all classes. In total, we considered 726 commits in our analysis, which contained 200,893 classes.

We extracted the TD items analyzing each snapshot with SonarQube's default rule set. 
To calculate fault-proneness, we determined fault-inducing and fault-fixing commits from the projects' Git history by applying the SZZ algorithm~\cite{SZZ}. 
The algorithm has four steps. The first step fetches the issues from a bug tracking system. All of the projects analyzed in this paper use Jira as their bug tracking system. The second step preprocesses the git log output, and the third identifies the bug-fixing commits. This is possible because the AFS policies require developers to report the fault-ID in the commit message of a fault-fixing commit. Finally, the last step identifies the fault-introducing commits using the data gathered in the previous steps.

\subsection{Data Analysis}
\label{DataAnalysis}

In order to answer our \textbf{RQs}, we investigated the differences between classes that are not affected by any TD items (clean classes) and classes affected by at least one TD item (dirty classes). This paper compares the change- and fault-proneness of the classes in these two groups. 

We calculated the class change- and fault-proneness adopting the same approach used by Palomba et al.~\cite{Palomba2018}.

We extracted the change logs from Git to identify the classes modified in each analyzed snapshot (one commit every 180 days). Then, we defined the change-proneness of a class $C_i$ in a commit $s_j$ as:

\begin{center}
{change-proneness}$_{C_i,s_j}$ = {$\#Changes(C_i)_{s_{j-1}\to s_j}$}
\end{center}

Where {$\#Changes(C_i)_{s_{j-1}\to s_j}$} is the number of changes made on $C_i$ by developers during the evolution of the system between the $s_j-1$ ’s and the $s_j$ ’s commit dates. 

SZZ provides the  list of fault-fixing commits and all the commits where a class has been modified to fix a specific fault. Therefore, we defined the fault-proneness of a class $c_i$ as the number of commits between snapshots $s_m$ and $s_n$ that fixed a fault in the program and altered the class $c_i$ in some way.

We calculated the normalized change- and fault-proneness for each class. The normalization was done by dividing the proneness value with the number of effective lines of code in the class. We defined an effective line of code as a non-empty line that does not start with "//", "/*", or "*". We also excluded lines that contained only an opening or closing curly bracket.

The results are presented using boxplots, which are a way of presenting the distribution of data by visualizing key values of the data. The plot consists of a box drawn from the 1$^{st}$ to the 3$^{rd}$ quartile and whiskers marking the minimum and maximum of the data. The line inside the box is the median. The minimum and maximum are drawn at 1.5*IQR (Inter-Quartile Range), and data points outside that range are not shown in the figure.

We also compared the distributions of the two groups using statistical tests. First, we determined whether the groups come from different distributions. This was done by means of the non-parametric Mann-Whitney test. The null hypothesis for the test is that when taking a random sample from two groups, the probability for the greater of the two samples to have been drawn from either of the groups is equal~\cite{Conover}. The null hypothesis was rejected and the distribution of the groups was considered statistically different if the p-value was smaller than $0.01$.
As Mann-Whitney does not convey any information about the magnitude of the difference between the groups, we used the Cliff's Delta effect size test. This is a non-parametric test meant for ordinal data. The results of the test were interpreted using guidelines provided by Grissom and Kim \cite{Grissom2005}. The effect size was considered negligible if $|d| < 0.100$, small if $0.100 \leq |d| < 0.330$, medium if $0.330 \leq |d| < 0.474$, and large if $|d| > 0.474$.

To answer \textbf{RQ1}, we compared the clean classes with all of the dirty classes, while for \textbf{RQ2}, we grouped the dirty classes based on the type of the different TD items and for \textbf{RQ3} by their level of severity. For each value of type and severity, we determined classes that were affected by at least one TD item with that type/severity value and compared that group with the clean classes. Note that one class can have several TD items and hence it can belong to several subgroups. For both \textbf{RQ2} and \textbf{RQ3} we used the same data, but in \textbf{RQ2} we did not care about the severity of the violated rule while on \textbf{RQ3} we did not care about the type.

Based on SonarQube's classification of TD items, we expected that classes containing TD items of the type \textit{Code Smell} should be more change-prone, while classes containing \textit{Bugs} should be more fault-prone. The analysis was done by grouping classes with a certain TD item and calculating the fault- and change-proneness of the classes in the group. This was done for each of the TD items and the results were visualized using boxplots. As with \textbf{RQ2} and \textbf{RQ3}, each class can contain several TD items and hence belong to several groups. Also, we did not inspect potential TD item combinations.
To investigate \textbf{RQ4}, we compared the type and severity assigned by SonarQube for each TD item with the actual fault-proneness, and change-proneness.

\subsection{Replicability}
\label{Replicability}
In order to allow our study to be replicated, we have published the complete raw data in the replication package\footnote{https://figshare.com/s/240a036f163759b1ec97}.

%% file: Sections/Results.tex
\section{Results}
\label{Results}
\subsection*{RQ1. Are classes affected by TD items more change- or fault-prone than non-affected ones?}

Out of 266 TD items monitored by SonarQube, 173 were detected in the analyzed projects. 

\begin{figure}[ht]
    \centering
    \includegraphics[width=0.7\columnwidth]{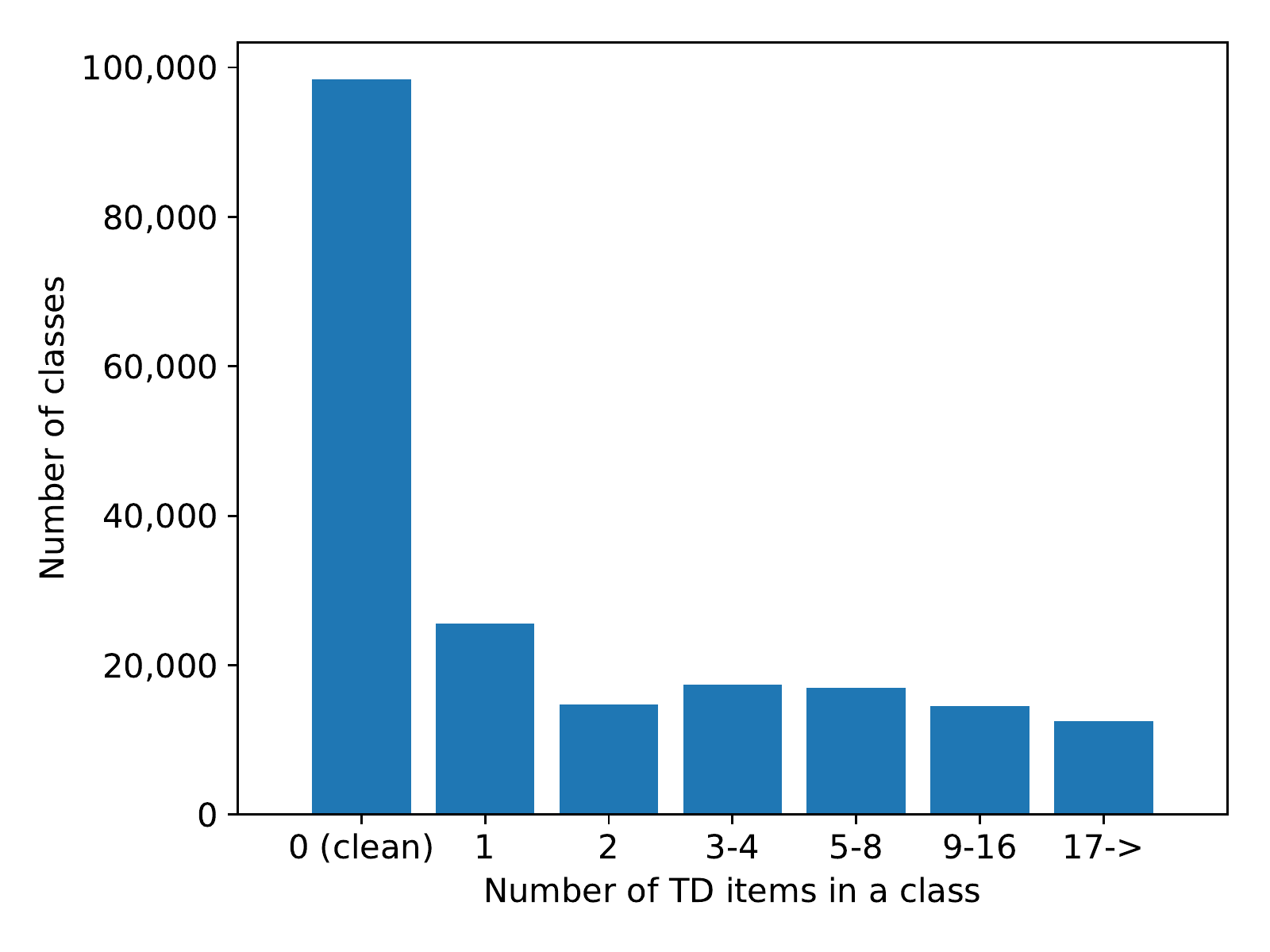}
    \caption{Number of classes with different numbers of TD items. (RQ1)}
    \label{fig:rq1_group_sizes}
\end{figure}

The analyzed commits contained 200,893 classes, of which 102,484 were affected by between 1 and 9,017 TD items. As can be seen from Figure~\ref{fig:rq1_group_sizes}, most of the classes were affected only by either zero or one TD item. However, the number of classes also dropped almost logarithmically as the number of TD items grew. This is why we grouped the results for the number of TD items in one class using the power of two as the limit for the number of TD items in a class. 

\begin{figure} []
    \centering 
    \begin{minipage}{0.4\textwidth}
    \centering
    \includegraphics[trim={0.85cm 0 0.05cm 0},clip,width=\columnwidth]{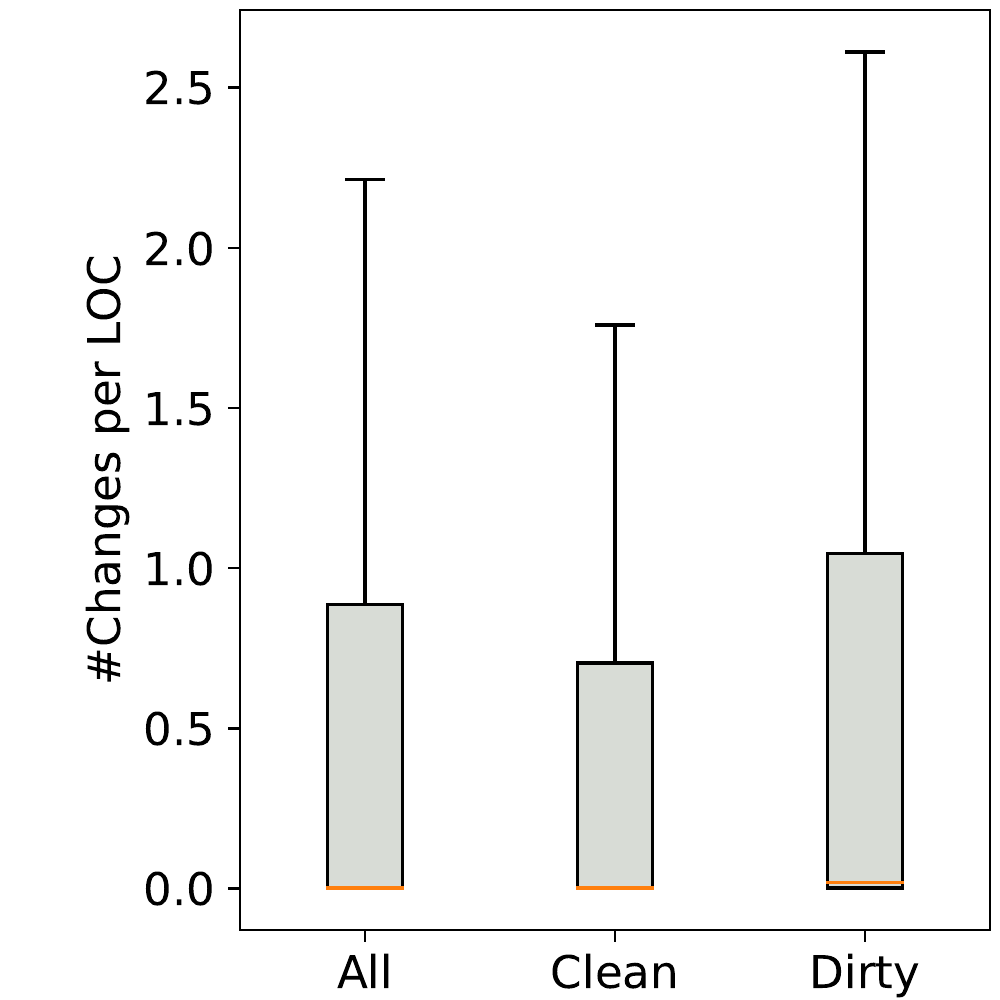}
    \caption{Change-proneness of classes affected by TD items (RQ1)}
    \label{fig:rq1_changeproneness_severity}
    \end{minipage}%
    ~\hspace{0.5cm}
    \begin{minipage}{0.6\textwidth}
        \centering
        \includegraphics[width=\columnwidth]{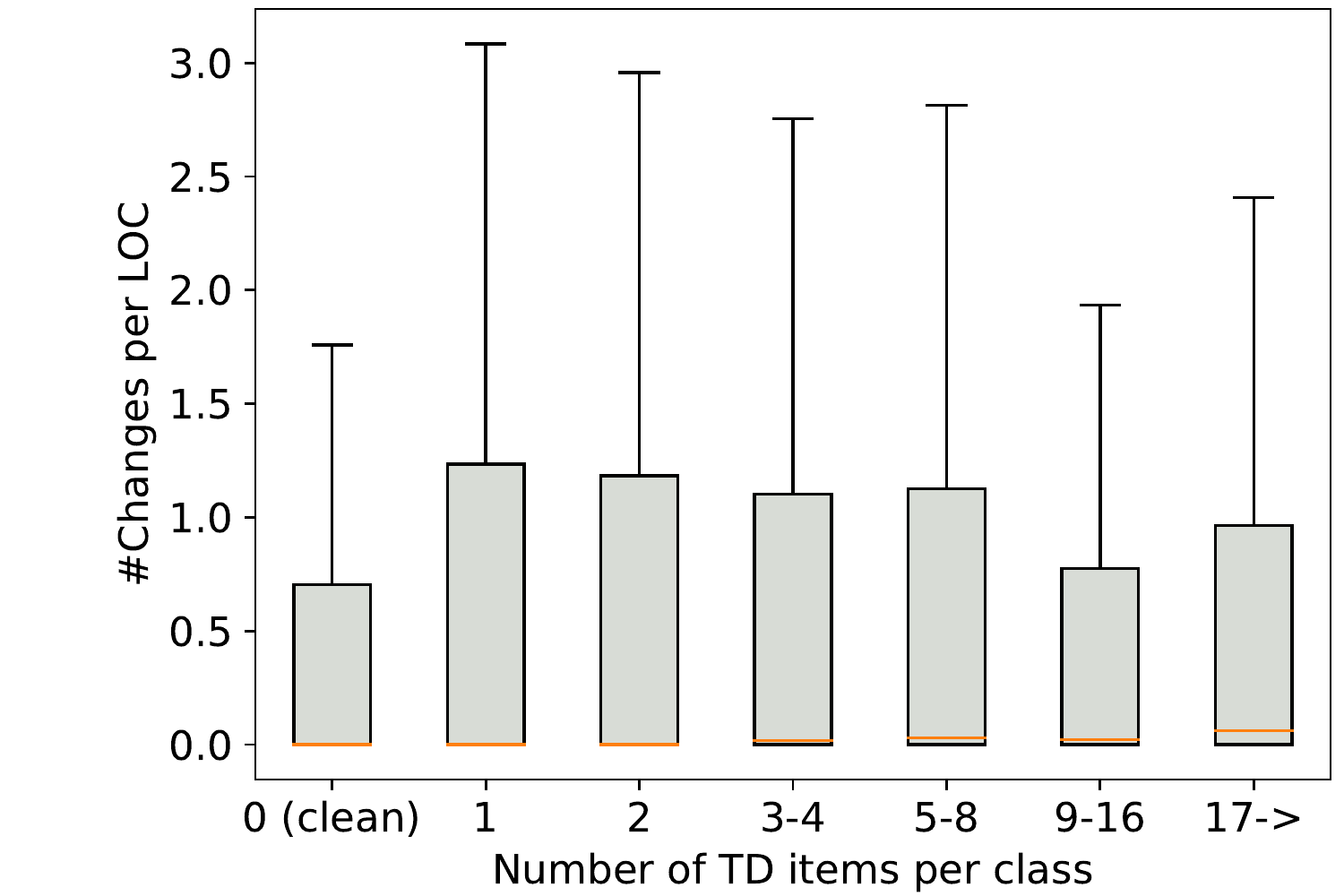}
        \caption{Change-proneness of classes affected by different numbers of TD items (RQ1)}
        \label{fig:rq1_change_number_of_TD_items}
    \end{minipage}%
\end{figure}

The distribution of the change-proneness of all of the classes in Figure~\ref{fig:rq1_changeproneness_severity} shows that the majority of the classes do not experience any changes and 75\% of the classes experience less than 0.89 changes per line of code. The figure also suggests that the differences in the distributions are minor between the clean and dirty classes. Dirty classes have a higher maximum (2.6), median (0.02), and Q3 (1.05), which is the third quartile containing 75 \% of the data. 

In order to identify the significance of the perceived differences between the clean and the dirty classes, we applied the Mann-Whitney and Cliff's Delta statistical tests. In terms of change-proneness, the p-value from the Mann-Whitney test was zero, which suggests that there is a statistically significant difference between the groups. The effect size was measured using Cliff’s delta. We measured a d-value of -0.06, which indicates a small difference in the distributions.

The fault-proneness of the classes is not visualized as the number of faults in the projects is so small, that also the maximum of the boxplot was zero. Thus, all of the faults were considered as outliers. However, when the statistical tests were run with the complete data, the p-value from the Mann-Whitney test was zero. This means there is a statistically significant difference between the two groups. However, the effect size was negligible, with d value of -0.005.

Moreover, we investigated the distributions of the change- and fault-proneness of classes affected by different numbers of TD items. We used the same groups as in Figure~\ref{fig:rq1_group_sizes}. 

The number of issues in a class does not seem to greatly impact the change-proneness (Figure~\ref{fig:rq1_change_number_of_TD_items}). The only slightly different group is the group with 9-16 issues as its Q3 is slightly less than for the other dirty groups. 

The results from the statistical tests confirm that the number of TD items in the class does not affect the change- or fault-proneness of the class (Table~\ref{tab:rq1_items_statistics}). Considering change-proneness, the Mann-Whitney test suggested that the distribution would differ for all groups. However, the Cliff's Delta test indicated that the differences are negligible for all groups except the one with 17 or more items, for which the difference was small. Thus, differentiating the dirty group into smaller subgroups did not change the previously presented result.

Once again, the fault-proneness is not visualized as the non-zero values were considered as outliers. In addition, while the statistical tests reveal that only the group with three or four TD items was found to be similar to the clean group, all of the effect sizes were found negligible.
 
\begin{table}[H]
    \footnotesize
    \centering
    \caption{Results from the Mann-Whitney (MW) and Cliff's Delta tests when comparing the group of clean classes with groups of classes affected by different numbers of TD items (RQ1)}
    \label{tab:rq1_items_statistics}
    \begin{tabular}{c|c|c|c|c}
    \hline
\multirow{2}{*}{\textbf{\#TD items per class}}  &  \multicolumn{2}{c|}{\textbf{change-proneness}} & \multicolumn{2}{c}{\textbf{fault-proneness}} \\ \cline{2-5}
     & \textbf{MW (p)} & \textbf{Cliff (d)} & \textbf{MW(p)} & \textbf{Cliff (d)} \\ \hline
         1 & 0.00 & -0.048 & 0.00 &  0.009 \\
         2 & 0.00 & -0.055 & 0.00 & 0.005 \\
         3-4 & 0.00 & -0.061 & 0.82 & -0.000 \\
         5-8 & 0.00 & -0.075 & 0.00 & -0.010 \\
         9-16 & 0.00 & -0.063 & 0.00 & -0.016 \\
         17$\rightarrow$ & 0.00 & -0.133 & 0.00 & -0.036 \\
         \hline
    \end{tabular}
\end{table}

\begin{figure}
    \begin{subfigure}[t]{0.5\textwidth}
        \includegraphics[width=\textwidth]{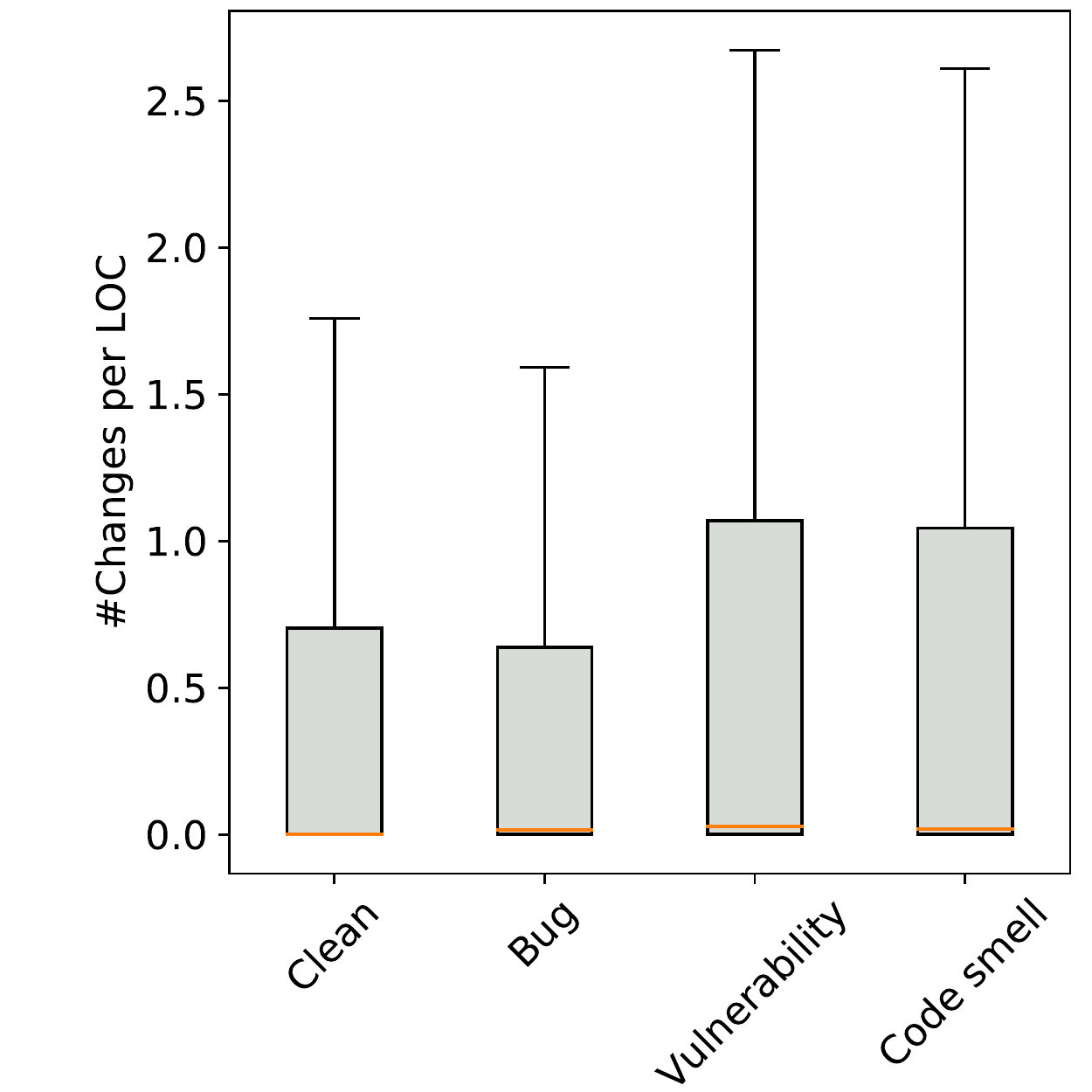}
        \caption{Change-proneness (type)}
        \label{fig:rq2_changeproneness_type}
    \end{subfigure}\hfill
    \begin{subfigure}[t]{0.5\textwidth}
        \includegraphics[width=\textwidth]{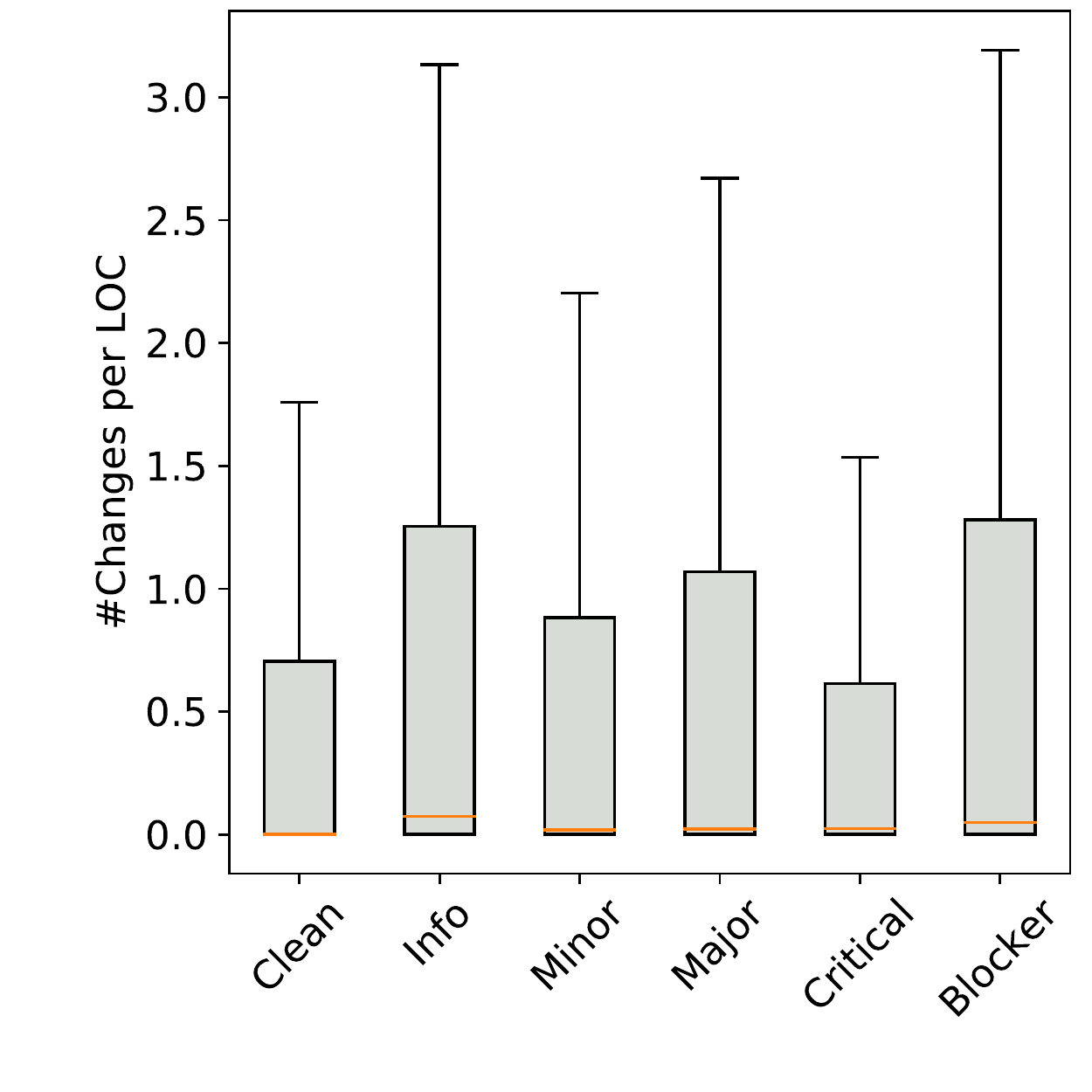}
        \caption{Change-proneness (severity)}
        \label{fig:rq2_changeproneness_severity}
    \end{subfigure}
    \caption{Change-proneness of classes affected by TD items considering type and severity (RQ2 and RQ3)} 
    \label{fig:rq2_change_fault_boxplots} 
\end{figure}

 However, the Cliff's Delta test indicated that the differences are negligible for all groups except the one with 17 or more items, for which the difference was small.
 In terms of fault-proneness, only the group with three or four TD items was found to be similar to the clean group. However, all of the effect sizes were determined to be negligible.

\vspace{2mm}
\hspace{-5mm}
\begin{tikzpicture}
\node [mybox] (box){%
\centering
\begin{minipage}{.95\textwidth}
\textbf{Summary of RQ1}\\
When inspecting six-month periods, the majority of the classes do not change and the rest of the classes have less than 2.5 changes per code line. Clean classes might be less change-prone than dirty classes, but the difference between the groups is small. When inspecting for fault-proneness, the code generally does not contain any faults and there is no difference between the clean and dirty classes. The number of TD items in a class does not remarkably affect the change- or fault-proneness.
\end{minipage}
};
\end{tikzpicture}%


\subsection*{RQ2. Are classes affected by TD items classified by SonarQube as different types more change- or fault-prone than non-affected ones?}

The diffuseness of the detected 173 TD rules grouped by type is reported in Table~\ref{tab:TDDiffuseness}.
We collected data regarding the number of classes affected by each Type and Severity of TD items (\textbf{\# affected classes}).
Moreover, we included the violated TD type recurrences (\textbf{\# rules}) and how many times they are violated (\textbf{\# introductions}).

The change-proneness of different types of dirty classes is provided in Figure~\ref{fig:rq2_changeproneness_type}. Fault-proneness is not visualized as the plot consists of only zeros.

Looking at the change-proneness of the different types, the distributions are divided in two groups. The most diffused types are \textit{Vulnerability} and \textit{Code Smell}, for which all of the key values are similar, with Q3 being approximately 1 and the maximum being 2.5. The less diffused groups are the \textit{Bug} type and the clean classes, which are similar to each other in terms of Q3 and maximum.

Moreover, the Mann-Whitney test suggested that regardless of the type, the distributions of the dirty groups would differ from the distribution of the clean group. However, the measured effect size was negligible for all of the types (Table~\ref{tab:rq2}).

Regarding the evaluation of fault-proneness, the distribution for the number of faults per code line in a class consists only of zeros and outliers for all of the inspected groups. Thus it is not visualized in the paper. Moreover, there do not appear to be any significant differences between the clean and the dirty groups. The Mann-Whitney test suggests that only the \textit{Bug} type does not have a statistically significant difference in the distribution, with a p-value of 0.07. For the other types, the p-value was less than 0.01. However, the Cliff's Delta test suggest that all of the effect sizes are negligible as the $|d|$ values are smaller than 0.1. 

\begin{table} []
\footnotesize
\centering
\caption{Diffuseness of detected TD items (RQ2 and RQ3)} 
\label{tab:TDDiffuseness} 
\begin{tabular}
{@{}p{1.7cm}|p{1.3cm}|p{1.4cm}|p{3.1cm}|p{2.7cm}@{}}
\hline 
\textbf{Type}& \textbf{Severity} & \textbf{\# Rules} & \textbf{\# Affected Classes} & \textbf{\# Introductions} \\\hline 
\multirow{6}{*}{Bug}	&	All	&	36	&	2,865	&	1,430	\\	
	&	Info	&	0	&	0	&	0	\\	
	&	Minor	&	0	&	0	&	0	\\	
	&	Major	&	8	&	1,324	&	377	\\	
	&	Critical	&	23	&	2,940	&	816	\\	
	&	Blocker	&	5	&	662	&	237	\\	\hline
\multirow{6}{*}{Code Smell}	&	All	&	130	&	70,822	&	132,173	\\	
	&	Info	&	2	&	12,281	&	5,387	\\	
	&	Minor	&	32	&	70,426	&	44,723	\\	
	&	Major	&	80	&	78,676	&	73,894	\\	
	&	Critical	&	14	&	19,636	&	7,556	\\	
	&	Blocker	&	2	&	1,655	&	613	\\	\hline
\multirow{6}{*}{Vulnerability}	&	All	&	7	&	3,556	&	2,241	\\	
	&	Info	&	0	&	0	&	0	\\	
	&	Minor	&	0	&	0	&	0	\\	
	&	Major	&	2	&	2,186	&	876	\\	
	&	Critical	&	5	&	3,490	&	1,365	\\	
	&	Blocker	&	0	&	0	&	0	\\	\hline
\textbf{Total}	&		&	173	&	102,484	&	135,844	\\	
\hline
\end{tabular}
\end{table}

\begin{table}[]
    \footnotesize
    \centering
    \caption{Results from the Mann-Whitney (MW) and Cliff's Delta tests when comparing the group of clean classes with groups of classes affected by TD items of different levels of severity and different types (RQ2 and RQ3)}
    \label{tab:rq2}
    \begin{tabular}{cc|c|c|c|c}
    \hline
\multicolumn{2}{c|}{\textbf{Severity and Type}}  &  \multicolumn{2}{c|}{\textbf{Change-proneness}} & \multicolumn{2}{c}{\textbf{Fault-proneness}} \\ \cline{3-6}
    & & \textbf{MW (p)} & \textbf{Cliff (d)} & \textbf{MW(p)} & \textbf{Cliff (d)} \\ \hline
    \multirow{5}{*}{Severity} & Info &  0.00 & -0.144  & 0.00 & -0.036 \\
         & Minor & 0.00 & -0.062 & 0.00 &  -0.009 \\
         & Major & 0.00 & -0.068 & 0.00 & 0.008 \\
         & Critical & 0.00 & -0.059 & 0.00 & -0.018 \\
         & Blocker & 0.00 & -0.101 & 0.00 & -0.066 \\
         \hline
         \multirow{3}{*}{Type} & Bug &  0.00  & -0.054 & 0.07 & -0.004 \\
         & Code Smell & 0.00 & -0.065 & 0.00 & -0.005 \\
         & Vulnerability & 0.00 & -0.072  & 0.00 & -0.022 \\
         \hline
    \end{tabular}
\end{table}

The results from the Mann-Whitney and Cliff's Delta tests are shown in Table~\ref{tab:rq2}. In terms of change-proneness, the Mann-Whitney test suggested that regardless of the type, the distributions of the dirty groups would differ from the distribution of the clean group. However, the measured effect size was negligible for all of the types.

Considering fault-proneness, the data is not visualized as the non-zero values were considered as outliers. The Mann-Whitney test suggests that only the \textit{Bug} type does not have a statistically significant difference in the distribution, with a p-value of 0.07. For the other types, the p-value was less than 0.00. However, the results from the Cliff's Delta test suggest that all of the effect sizes are negligible as the $|d|$ values are smaller than 0.1. 

\vspace{2mm}
\hspace{-5mm}
\begin{tikzpicture}
\node [mybox] (box){%
\centering
\begin{minipage}{.95\textwidth}
\textbf{Summary of RQ2}\\
Considering the \textbf{type} of different TD items, the types \textit{Vulnerability} and \textit{Code Smell} seem to be slightly more change-prone than the clean classes, but the differences are negligible. We did not find any significant differences regarding the fault-proneness of the classes. 
\end{minipage}
};
\end{tikzpicture}%

\subsection*{RQ3. Are classes affected by TD items classified by SonarQube with different levels of severity more change- or fault-prone than non-affected ones?}

Table~\ref{tab:TDDiffuseness} reports the diffuseness of the detected 173 TD rules grouped by severity. 

The change-proneness of the dirty classes regarding different types is provided in Figure~\ref{fig:rq2_changeproneness_severity}. The most diffused levels are the least and the most severe levels \textit{Info} and \textit{Blocker}. Both of these groups have medians greater than zero, meaning most of the data does not consist of zeros. The median for \textit{Info} is 0.07 and for \textit{Blocker} it is 0.05, while their maximums are more than three changes per line of code and the Q3s are around 1.2. The least diffused level is \textit{Critical}, while the levels \textit{Major} and \textit{Minor} are in the between.

The results from the Mann-Whitney and Cliff's Delta tests for the different severity levels are given in Table~\ref{tab:rq2}. The distribution of the change-proneness of all the groups was found to be different than that for the clean group. However, the measured effect size was negligible for the severity levels \textit{Minor}, \textit{Major}, and \textit{Critical}, and small for the levels \textit{Info} and \textit{Blocker}. The results from the statistical tests confirmed the visual results from the boxplots, namely, that there are no significant differences between the clean classes and the different values of severity.

Considering the fault-proneness of the different severity levels, the results are similar to the fault-proneness of the different type values. The Mann-Whitney test suggests that the distributions would differ for all levels, but when the effect size was measured, it was found to be negligible for every level.

\vspace{2mm}
\hspace{-5mm}
\begin{tikzpicture}
\node [mybox] (box){%
\centering
\begin{minipage}{0.95\textwidth}
\textbf{Summary of RQ3}\\
Regarding \textbf{severity}, the dirty classes are not significantly more change-prone than the clean classes either. We did not find any significant differences regarding the fault-proneness of the classes. 
\end{minipage}
};
\end{tikzpicture}%

\subsection*{RQ4. How well are SonarQube's TD rules classified?}
\begin{figure}
    \centering
    \includegraphics[width=
    \textwidth]{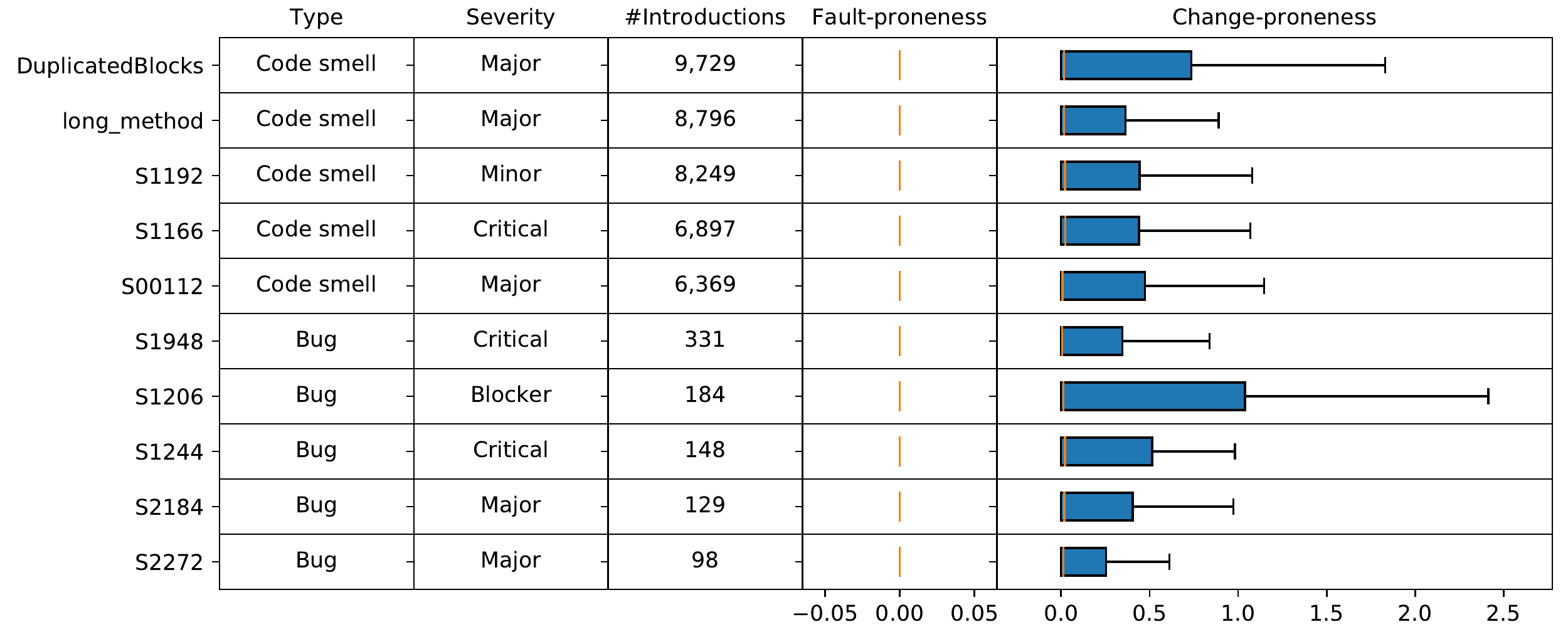}
    \caption{The change-proneness of classes infected by TD items defined as \textit{Code Smell} and the fault-proneness of classes infected by TD items defined as \textit{Bug} (the five most frequently introduced TD items) - (RQ4)}
    \label{fig:RQ4}
\end{figure}

In order to confirm the classification of SonarQube's TD items, we inspected the change- and fault-proneness of single TD items. For reasons of space, Figure~\ref{fig:RQ4} shows only the distribution of the five most frequently introduced TD items for the types \textit{Bug} and \textit{Code Smell}. The complete figure is available in the replicated package (Section~\ref{Replicability}). The results are unexpected, since when we look at the \textit{Bug} type, we can see that none of the TD items are fault-prone, but they are change-prone. This confirms the results obtained in the previous RQs. Regarding \textit{Code Smells}, TD items are change-prone even if their assigned level of severity is never confirmed. (The most relevant examples are UselessImportanCheck and S1166). 
We considered all the TD items classified as \textit{Bug} and \textit{Code Smell} with the assigned level of severity affecting the analyzed projects.




\textbf{Bug}. All the TD items classified as Bug have no impact on maintainability since they turned out not to be fault-prone. This confirmed the results obtained for the previous RQs. Instead, and unexpectedly in addition, they appear to have a clear impact, which on some occasions is just as relevant, on code stability. For example, S1860, OverridenCallSuperFinalizeCheck, has a moderate impact, while S1175 and S2066 have a higher impact.



\textbf{Code Smells}. Unexpectedly, two code smells (RightCurlyBraceStartLineCheck and S1171) were also found to have a relevant impact on fault-proneness. 
Regarding the level of severity, we found only two code smells classified as \textit{Blocker} (S2442 and S1181).
We found 13 code smells with the severity level \textit{Critical}. Unexpectedly, only two code smells (S2178 and S1994) can confirm the assigned severity level. One code smell, S2447, seems to have a lower severity level than the assigned one.
Out of 80 code smells with the severity level \textit{Major}, we found only nine that confirm the severity level, while the level of severity assigned to the others by SonarQube was found to be overestimated. 
Moreover, we found an underestimation of four code smells with the severity level \textit{Minor } (UnlessImportanCheck, S00119, RightCurlyBraceStartLineCheck, and S1195). Actually, they have a higher impact on code stability.

\vspace{2mm}
\hspace{-5mm}
\begin{tikzpicture}
\node [mybox] (box){%
\centering
\begin{minipage}{.95\textwidth}
\textbf{Summary of RQ4}\\
The change-proneness of the classes affected by TD items of the type \textit{Code Smell} differs notably, even though the majority of the affected classes do not change. TD items of the type \textit{Bug} are not more fault-prone, as the roneness does not appear to be dependent on the type of the TD item and the assigned level of severity. 
\end{minipage}
};
\end{tikzpicture}%

%% file: Sections/Discussion.tex
\section{Discussion}
\label{Discussion}

In this Section, we will discuss the results obtained according to the RQs and present possible practical implications from our research.

\textbf{Answers to Research Questions}. The analysis of the evolution of 33 Java projects showed that, in general, the number of TD Items in a class does not remarkably affect the change- or fault-proneness. Clean classes (classes not affected by TD Items) might be less change prone than dirty classes, but the difference between the groups is small. 
Moreover, when inspecting for fault-proneness, the code generally does not contain any faults and there is no difference between the clean and dirty classes.


Out of 266 TD items detected by SonarQube, we retrieved 173 in the analyzed projects: 36 \textit{''Bugs''}, 130 \textit{''Code Smells''}, and 7 \textit{''Vulnerabilities''}.

Taking into account TD Items classified as \textit{Bug} (supposed to increase the fault-proneness) only one increases the fault-proneness. Unexpectedly, all the remaining \textit{Bugs} resulted to slightly increase the change-proneness instead. 
As expected, all the 130 TD items classified as \textit{Code Smell} affect change-proneness, even if their impact on the change-proneness is very low.
Moreover, also the seven TD items classified as \textit{Vulnerability}  have a low effect on change-proneness.


However, the change- and fault-proneness of the vast majority of TD items (more than 70\%)
does not always increase together with the severity level assigned by SonarQube.

\textbf{Implications}. SonarQube recommends manual customization of their set of rules instead of using the out-of-the-box rule set. 
However, as reported by \cite{SaarimakiTechDebt2019}, querying the SonarQube public instance APIs\footnote{https://docs.sonarqube.org/display/DEV/API+Basics}, we can see that more than 98\% of the public projects (14,732 projects up to 14,957) use the "sonar way" rule set, mainly because developers have no experience with customizing nor understand which rules are more change  or fault prone. 

Our results are similar to  Tollin et al.~\cite{Tollin2017}, even if in our case the effect of TD items on change-proneness is very low. Tollin et al. found an increase in change-proneness in classes affected by TD items in two industrial projects written in C\# and Javascript. However, they adopted C\# and Javascript rules, which are different from the Java rules. The difference in the results regarding change-proneness could either be due to the different projects (33 open-source Java projects) or to the different rules defined for Java. 

The main implication for practitioners is that they should carefully select the rules to consider when using SonarQube, especially if they plan to invest effort to reduce the change- or fault-proneness. We recommend that practitioners should apply a similar approach as the one we adopted, performing a historical analysis of their project and classifying the actual change- and fault-proneness of their code, instead of relying on their perception of what could be fault or change prone. 
Researchers should continue to study this topic in more depth and help both practitioners and tool providers to understand the actual harmfulness of TD items, but should also propose automated tools for performing historical analyses in order to automatically identify the harmfulness of TD items. 

%% file: Sections/Threats.tex
\section{Threats to Validity}
\label{Threats}
In this Section, we will introduce the threats to validity, following the structure suggested by Yin~\cite{YinCaseStudies2009}.

\textbf{Construct Validity}. This threat concerns the relationship between theory and observation. 
\textbf{Limitations}. We adopted the measures detected by SonarQube, since our goal was to analyze the diffuseness of TD items  in  software  systems  and  to  assess  their  impact  on  the change- and fault-proneness of the code, considering also the type of TD items and their severity.
We are aware that the SonarQube detection accuracy of some rules might not be perfect, but we tried to replicated the same conditions adopted by practitioners when using ithe same tool.  
Unfortunately, several projects in our dataset do not tag the releases. Therefore, we evaluated the change- and fault-proneness of classes as the number of changes and bug fixes a class was subject to in a period of six months. We are aware that using releases could have been more accurate for faults. However, as Palomba et al.~\cite{Palomba2018} highlighted, the usage of releases has the threats that time between releases is different and the number of commits and changes are not directly comparable. Unfortunately Git does not provide explicit tags for several projects in our dataset. 
We relied on the SZZ algorithm~\cite{SZZ} to classify fault-inducing commits. We are aware that SZZ provides a rough approximation of the commits inducing a fix because of Git line-based limitations of Git and because a fault can be fixed also modifying a different set of lines than the inducing ones. 
Moreover, we cannot exclude misclassification of Jira issues (e.g., a new feature classified as bug).
As for the data analysis, we normalized change- and fault-proneness per class using LOC. As alternative other measures such as complexity could have been used.


\textbf{Internal Validity}. This threat concerns internal factors of the study that may have affected the results.
Some issues detected by SonarQube were duplicated, reporting the issue violated in the same class and in the same position but with different resolution times. We are aware of this fact, but we did not remove such issues from the analysis since we wanted to report the results without modifying the output provided by SonarQube. We are aware that we cannot claim  a direct cause-effect relationship between the presence of a TD items and the fault- and change-proneness of classes, that can be influence by other factors. We are also aware that classes with different roles (e.g., classes controlling the business logic) can be more frequently modified than others. 

\textbf{External Validity.} This threat concerns the generalizability of the results. We selected 33 projects from the Apache Software Foundation, which incubates only certain systems that follow specific and strict quality rules. Our case study was not based only on one application domain. This was avoided since we aimed to find general mathematical models for the prediction of the number of bugs in a system. Choosing only one or a very small number of application domains could have been an indication of the non-generality of our study, as only prediction models from the selected application domain would have been chosen. The selected projects stem from a very large set of application domains, ranging from external libraries, frameworks, and web utilities to large computational infrastructures. The application domain was not an important criterion for the selection of the projects to be analyzed, but in any case we tried to balance the selection and pick systems from as many contexts as possible. We are considering only open source projects, and we cannot speculated on industrial projects. Moreover, we only considered Java projects due to the limitation of the tools used (SonarQube provides a different set of TD items for each language) and results would have not been directly comparable. 

\textbf{Reliability}. This threat concerns the relationship between the treatment and the outcome. We do not exclude the possibility that other statistical or machine learning approaches, such as Deep Learning, might have yielded similar or even better accuracy than our modeling approach.

%% file: Sections/Conclusion.tex
\section{Conclusion}
\label{Conclusion}
In this paper, we studied the impact of TD items on change- and fault-proneness, considering also the type and severity, based on  33 Java systems from the Apache Software Foundation. We analyzed nearly 726 commits containing 27K faults and 12 million changes. The projects were infected by 173 SonarQube TD items violated more than 95K times in more than 200K classes analyzed. 


Our results revealed that dirty classes might be more prone to change than classes not affected by TD items. However, the difference between the clean and dirty groups was found to be at most small regardless of the \textbf{type} and \textbf{severity}. When considering the fault-proneness of the classes, no significant differences were found between the clean classes and the groups with dirty classes.
As for SonarQube classification of TD items, all the TD items, including all the \textit{Bugs}, \textit{Code Smells} and \textit{Vulnerabilities} have a statistically significant, but very small effect on change-proneness. 
Only one out of 36 TD items classified as \textit{Bug} (supposed to increase the fault-proneness) has a very limited effect on fault-proneness. 

Our study shows that SonarQube could be useful and TD items should be monitored by developers since all of them are related to maintainability aspects such as change-proneness. 
Despite our results show that the impact on change-proneness of TD items is very low, monitoring projects with SonarQube would still help to write cleaner code and to slightly reduce change-proneness.  As recommended by SonarQube, we would not recommend to invest time refactoring TD items if the goal is to reduce change- or fault-proneness, instead we would recommend not to write new code containing TD items. 
The result of this work can be useful for practitioners and help them to understand how to prioritize TD items they should refactor. It can also be used by researchers to bridge the missing gaps, and it supports companies and tool vendors in identifying TD as accurately as possible. 

As regards future work, we plan to further investigate the harmfulness of SonarQube TD items, also comparing them with other types of technical debt, including architectural and documentation debt. 
We are planning to replicate this work, adopting different analysis techniques, including machine learning. Moreover, we also plan to conduct a case study with practitioners to understand the perceived harmfulness of TD items in the code.